\documentclass[twocolumn,showpacs,preprintnumbers,amsmath,amssymb,aps]{revtex4}

\usepackage{graphicx}
\usepackage{dcolumn}
\usepackage{bm}

\begin{document}

\title{Overlapping resonances in nuclei coupling to the atomic shells}

\author{Srinivas K. Arigapudi\footnote{On leave from Indian Institute of Technology, New Delhi 110 016, India}}\email{sinu.iitd@gmail.com}
\affiliation{Max-Planck-Institut f\"ur Kernphysik,
Saupfercheckweg 1, D-69117 Heidelberg, Germany}

\author{Adriana P\'alffy}\email{palffy@mpi-hd.mpg.de}
\affiliation{Max-Planck-Institut f\"ur Kernphysik,
Saupfercheckweg 1, D-69117 Heidelberg, Germany}

\date{\today}

\begin{abstract}
The resonant process of nuclear excitation by electron transition (NEET) in highly charged ions is investigated. In NEET, 
 a bound  electronic decay transition  occurs with the simultaneous excitation of the nucleus, provided that the energies of the atomic and nuclear transition match. 
By varying the atomic charge state, the atomic transition energy can be tuned to a better match of the nuclear transition energy. We propose a new way to create the atomic hole in highly charged ions by dielectronic capture of a free electron. In order to derive the cross section for the three-step process composed by dielectronic capture, nuclear excitation by electron transition, and the subsequent nuclear decay, a Feshbach projection operator formalism is developed. With the help of this formalism, the resonances in nuclei coupling to the atomic shell and the possible interference between several direct and resonant electronic and nuclear processes such as radiative recombination, dielectronic recombination and nuclear excitation by electron capture are described.

\end{abstract}

\maketitle


\section{Introduction}

The increasing precision in atomic physics experiments ignited in the last decades the nuclear odyssey from the simplified picture of a point-like charge to its real size and properties. Different nuclear masses,  charge distributions or spins  have an effect on the electronic  transition energies, and the small energy corrections or splittings observed in experiments can give in turn information about the nucleus. Atomic spectroscopy,  at present at an unrivaled level of precision, can thus be used to determine nuclear parameters which are otherwise hardly accessible via nuclear physics experiments \cite{contemp2010}. 

On the other hand, the borderline between atomic and nuclear physics is also the scene for several nuclear processes that directly involve atomic electrons.  For instance,  nucleus and electron can interact via the electromagnetic field and undergo transitions simultaneously.   A nucleus in an excited state that  cannot decay  radiatively  transfers its energy via the radiation field to one of the atomic electrons which leaves the atom in the process of internal conversion (IC). The inverse process of IC, which might occur in highly-charged ions (HCI), is known as nuclear excitation by electron capture (NEEC) \cite{palffy2006}. In conjunction with HCI, atomic physics experiments involving these processes at the borderline between atomic and nuclear physics also open the possibility to explore properties of exotic nuclei \cite{contemp2010}.

IC and NEEC also have  more exotic siblings, the {\it bound} internal conversion (BIC) and its inverse process NEET. Bound IC is a resonant nuclear decay channel which may occur if the nuclear excitation energy is not enough to ionize the atomic electron, but can induce a transition between two bound electronic states.  NEET is  the simultaneous excitation of the nucleus  during an atomic decay transition, provided the energies of the two transitions match. There is only a limited number of nuclei in which such a match of the atomic and nuclear transition energies exist, see for instance the list of candidates for NEET in Ref.~\cite{NEET_list}. 
Although it is difficult to find  systems that fulfill the  resonance condition,  NEET has already been observed experimentally  \cite{Kishimoto2000}, in the same year in which direct evidence of BIC \cite{Carreyre} has been reported.  As nuclear excitation mechanisms, both  NEEC and NEET are  expected to allow the study of atomic vacancy effects on nuclear level population and lifetimes.

The NEET nuclear excitation mechanism has been first proposed by Morita back in 1973 \cite{Morita1973} as means  for $^{235}\mathrm{U}$ separation. This first theoretical work was followed only months later by a first claimed experimental observation \cite{Otozai1973} in $^{189}\mathrm{Os}$. However, these early claims of success, as well as the follow-up published experimental results \cite{Saito1980,Fujioka1984,Otozai1978,Shinohara1987} were riddled with controversy. While theorists continued arguing about the true magnitude of the NEET probability \cite{Pisk1989,Tkalya1992,Ho1993}, apparently overestimated in the original proposal \cite{Morita1973}, many  studies dedicated to NEET in $^{235}\mathrm{U}$ in plasma environments were performed \cite{Chemin1999,Morel2004PRA}. Other theoretical predictions confirming the important role that  nuclear excitation mechanisms such as NEEC and NEET may have in dense astrophysical plasmas were also more recently published \cite{Gosselin2007,Morel2010}.

The most recent and reliable experimental results on NEET have been performed on $^{197}\mathrm{Au}$ \cite{Kishimoto2000,Kishimoto2006}, $^{189}\mathrm{Os}$ \cite{Ahmad2000,Aoki2001} and $^{235}\mathrm{U}$ \cite{Claverie2004}. 
The NEET probability for the 77.351 keV $^{197}\mathrm{Au}$ nuclear transition that corresponds to an atomic  $K\rightarrow M$ transition in neutral atoms was determined to be $(5.0\pm 0.6)\times 10^{-8}$ \cite{Kishimoto2000}. In the case of the 69.5 keV nuclear transition in $^{189}\mathrm{Os}$, only an upper limit of $4.1\times 10^{-10}$ for the NEET probability could be deduced from experiments \cite{Aoki2001}, while for $^{235}\mathrm{U}$ the occurrence of NEET could not be observed. 

From the theoretical point of view, a number of sometimes contradictory studies have been considering the NEET mechanism in the last decades, among which we mention the nonrelativistic self-consistent description for x-ray, Auger decay and NEET \cite{Ho1993}, the relativistic analysis of NEET  \cite{Tkalya1992} and a formalism developed for NEET in plasmas at local thermodynamical equilibrium \cite{Morel2004PRA}. A comparison between theoretical results seeking to clarify all anomalies and discrepancies is presented in Ref.~\cite{Harston2001}. Furthermore, the efficiency of NEET as nuclear excitation mechanism for isomer depletion has been studied for a number of special cases \cite{TkalyaHf,KarpeshinTh} and was also compared or considered along with NEEC or photoexcitation \cite{Zadernovsky2002,Morel2010,Gosselin2007,Chemin1999}.

The purpose of the present work is manifold. First, we proceed by arguing on behalf of NEET experiments with HCI. While atomic physics experiments with HCI continuously gained in precision, until now all the performed NEET experiments envisaged neutral atoms irradiated by synchrotron radiation light to create the necessary inner-shell electronic hole. Furthermore, although in the theoretical studies of NEET in plasmas the general importance of the ionic charge state and the environment has been already mentioned, we emphasize here in particular the possibility that HCI offer to ``tune'' the atomic transition energy. A number of elements  that present a mismatch on the order of 1~keV between the nuclear and atomic transition energies ($ E_n$ and $ E_a$, respectively) for neutral atoms   offer the possibility to reduce the energy difference $(E_n- E_a)$ by at least one order of magnitude for other charged states. As it will be shown in this work, such an improved match of the atomic and nuclear transition energies can determine an increase of three orders of magnitude in the NEET probability. Considering the small number of  energy match cases that can be found in nature, the partial tunability of the atomic energy brings additional NEET candidate elements on the list and is most welcome. 

\begin{figure}[h]
  \centering
  \includegraphics*[width=8.7cm]
             {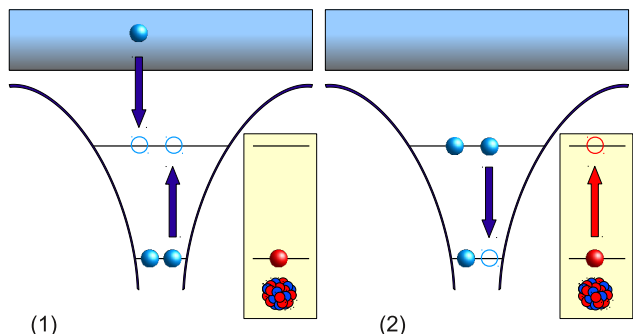}
  \caption{ (1) The dielectronic capture: a free electron recombines with the simultaneous excitation of a bound one. (2) NEET: electronic and nuclear transitions  occur simultaneously. The nuclear levels are depicted in the yellow panels.}
  \label{figure1}
\end{figure}

Next, starting from the premises of NEET in HCI, we consider a new mechanism for production of the electronic hole, namely dielectronic capture (DC). In the resonant process of DC, a free electron recombines into a HCI with the simultaneous excitation of a bound electron, which leaves a hole in the previously occupied bound shell. DC followed by x-ray decay of the excited electrons is known as dielectronic recombination (DR), an atomic physics process which is used in high-precision experiments with HCI to measure isotope shifts and provide complementary data to nuclear radii obtained via other methods \cite{Brandau2008}.
 The hole created by DC can be occupied via electronic decay that occurs simultaneously with the excitation of the nucleus, i.e., NEET. This two-step process is presented in Fig.~\ref{figure1}. 

By tuning the energy of the continuum electron one can precisely control which hole is created and which electronic transitions are driven. Furthermore, by using different types of DC, such as $KLL$-, $KLM$- or $KLN$-DC, (the notation specifies the involved atomic shells)  one can make use of satellite electrons in higher-$n$ shells to fine-tune the atomic transition energies. As a further advantage, 
depending on the optimal charge state to minimize the energy mismatch $(E_n- E_a)$, one can envisage a situation like in Fig.~\ref{figure1} in which the electronic hole can only be filled by an atomic transition that has the right energy to couple to the nuclear excitation. In contrast, a hole created by photoionization of an inner shell in a neutral atom has many decay channels corresponding to a multitude of electronic transitions. Furthermore,  most of the nuclear transitions that can couple to the atomic shell are dipole-forbidden, and the one  atomic transition which corresponds to NEET ought to share this multipolarity. In this case, the NEET atomic transition is  strongly suppressed in comparison with other electric dipole ($E1$) radiative or Auger decay channels of the atomic hole.
 A substantial gain can be thus achieved in a simple manner by using DC in HCI as first step previous to NEET in a scenario like the one presented in Fig.~\ref{figure1}. This scenario may also be relevant for dense plasmas, where DR is the dominant electron recombination mechanism.

We develop here a theoretical formalism that can treat the complex three-step process consisting of DC, NEET and the subsequent decay of the excited nuclear state. In order to derive the
cross section formula for this process we have extended the Feshbach projector
formalism developed and used for DR \cite{Zimmermann} and NEEC \cite{palffy2006} to account for the
interaction of electronic and nuclear degrees of freedom. The electric and magnetic electron-nucleus interactions are considered explicitly. The
dynamics of electrons is governed by the Dirac equation as required in the case of high-$Z$ elements. 

Our formalism allows us to also take into account interference processes that may occur in the sequence of resonant processes. We present here a case study for $^{237}\mathrm{Np}$ for which all possible interferences between radiative recombination (RR), DR, NEEC followed by gamma decay and the three-step NEET process under consideration are taken into account. The  interference between the channels of RR and DR  with the sequence of DC, NEET and radiative decay of the excited nuclear state  turns out to be rather weak,
but is still two to three orders of magnitude larger in cross section than the NEET process itself.  Reasons for the small interference cross sections are the very different time scales of the two processes, since the nuclear excitation needs a long time to occur as compared with the radiative recombination of a free electron into a HCI. This is not the case when considering interference between DC followed by NEET and NEEC, where both  channels start from the same initial state with an electron in the continuum  and have as final state an excited nucleus. However, depending on the ionic configuration and on the recombination state for DC, this interference channel may not always be present. We find that also the magnitude of this interference term between the two pathways involving nuclear excitation mechanisms is the same as the ones involving the atomic processes RR and DR. At the resonance energy,   NEEC turns out to be the dominant nuclear excitation mechanism and  acts as an enhancement of the nuclear excitation probability.


We would like to point out that, while for NEET the atomic transition matching the nuclear excitation is unique, NEEC has an additional degree of freedom for the electron capture into a bound shell as 
continuum electron energy and capture orbital can be varied.  From this respect a comparison between the two processes needs to specify the particular conditions, for instance, the ionization degree and the capture orbital for NEEC. This excludes a too general argument or criterion for the comparison between NEEC and NEET. In the framework of the process sequence that we consider---dielectron capture  in a highly charged ion followed by NEET and gamma decay---a unique choice for the NEEC channel is naturally arising when investigating the processes that can interfere. Previous results  \cite{Gosselin2007} considering the temperature-dependent NEEC and NEET rates in dense plasmas have shown that for lower ionization degrees, NEET is more efficient than the  possible NEEC channels. This feature was attributed to the better overlap between the bound electronic wave functions.  Here, we show that for few-electron high-$Z$ HCI this is not the case and NEEC may be the most significant contribution.

This paper is structured as follows. We start by developing the Feshbach formalism to account for a three step process consisting of DR followed by NEET and subsequent nuclear decay in Section~\ref{projform}. The specific transition rates are identified in the cross section and their expressions are reviewed in Section~\ref{rates}, followed by Section \ref{interf} where the interference terms are deduced and discussed. A numerical case study is presented in  Section~\ref{results}. The paper ends with a summary and outlook. Atomic units have
been used unless otherwise mentioned.


\section{\label{projform} Feshbach projection formalism}

In this section we derive the total cross section formula for a three-step process
consisting of DR, NEET and the decay of the excited nucleus. Without limiting too 
much the generality of our argument, we will treat here the radiative decay of
the excited nuclear state, denoted in the following by $\gamma$ decay. The case of nuclear excitation 
followed by IC can be treated in a similar manner, as it was shown for NEEC in Ref.~\cite{nres}.
 For simplicity we consider  the case  where the three relevant transitions energies (of the captured
 and bound electrons and of the nucleus) are equal. 
This situation is also relevant for
investigating the interference between DR followed by NEET and NEEC.
Furthermore, we treat the case in which the decay of the electronic hole occurs to the ground state, such that no further electronic transitions are possible.
The theory can easily be extended to account for an arbitrary DC electronic configuration.

\subsection{Decomposition of the Fock space by means of projection operators \label{subspaces}}
 The initial state $\psi_i$ of the system consisting of the nucleus in its ground state, the free electron, the bound electrons ground state configuration and the vacuum state of the electromagnetic field is given by 
\begin{equation}
|\psi_i\rangle  =  |I_gM_g, \alpha_i J_iM_i\vec{p}m_s,0\rangle\, .
\end{equation}
Here, $\vec{p}$ is the asymptotic momentum of the free electron and $m_s$ is its spin projection. The nuclear ground state  is denoted by the total angular momentum $I_g$ and its projection $M_g$. The bound electronic ground state consisting of $n$ electrons is denoted here by the total angular momentum of the electronic configuration $J_i$ and its projection $M_i$ and any other quantum numbers $\alpha_i$. 

In the first step DC takes place and the new state denoted here as first intermediate state $|\psi_{d_1}\rangle$ has the form 
\begin{equation}
|\psi_{d_1}\rangle = |I_gM_g,\alpha_{1}J_{1}M_{1},0\rangle \, ,
\end{equation}
where with $J_{1}$ and $M_{1}$ we have denoted  the total electronic angular momentum and its projection on the quantization axis, respectively, of the new electronic configuration of $n+1$ electrons,  and $\alpha_{1}$ covers again any other quantum numbers of this state. The nucleus is still in its ground state and no photon has been emitted. 

In the second step NEET occurs and the system reaches the second intermediate state $\psi_{d_2}$ of the form  
\begin{equation}
|\psi_{d_2}\rangle = | I_eM_e,\alpha_{2}J_{2}M_{2},0\rangle\, .
\end{equation}
The $n+1$ electrons form now a different configuration with total angular momentum $J_{2}$ and corresponding projection $M_{2}$, and the nuclear excited state  is described by the quantum numbers $I_e$ and $M_e$. The excited state of the nucleus  decays radiatively back to the ground state and the final state $\psi_f$ is given by 
\begin{equation}
|\psi_f\rangle = |I_gM_f,\alpha_{2}J_{2}M_{2},\lambda kLM\rangle\, .
\end{equation}
The emitted photon has the wave number $k$, the total angular momentum $L$ and  projection $M$. Furthermore, $\lambda$ stands for electric $(E)$ or magnetic $(M)$ spherical waves. The photonic state can be written as 
$|\lambda kLM\rangle = a^{\dagger}_{\lambda kLM}|0\rangle$, where $a^{\dagger}_{\lambda kLM}$ is the photon creation operator. The corresponding conjugate annihilation operator is denoted by $a_{\lambda kLM}$.

We now separate the Hilbert space into four possible subspaces 
\begin{itemize}
\item 
P -- states with a continuum electron, 
\item
$Q_1$ -- states with bound excited electrons and ground state nucleus, 
\item
$Q_2$ -- states with excited nuclei and ground state electrons (since we do not consider here the situation when after NEET further deexcitations of the electronic shell may occur),
\item 
R -- states with photon.
\end{itemize}

We subsequently introduce operators projecting onto individual subspaces. Characterizing the state of the electron in the positive part of the continuous spectrum by the energy $\epsilon$ rather than the momentum of the free electron, we write the projector P belonging to the first type of subspace as 
\begin{equation}
 P = \int d\epsilon \sum_{\beta}|\beta\epsilon\rangle\langle\beta\epsilon| \, . 
\end{equation}
For brevity we introduce the multi-index $\beta$ to stand for all discrete quantum numbers of the total system. The projection operator of the subspace spanned by intermediate states belonging to of the type $|\psi_{d_1}\rangle$ is written as 
\begin{equation}
Q_1 = \sum_{q_1}|q_1\rangle\langle q_1| \, , 
\end{equation}
with the cumulative index $q_1$ introduced to summarize all discrete quantum numbers describing the excited electron and the ground state nucleus. The projection operator of the subspace spanned by intermediate states of the type $|\psi_{d_2}\rangle$ is written as 
\begin{equation}
Q_2 = \sum_{q_2}|q_2\rangle\langle q_2| \, ,
\label{Q2spectral}
\end{equation}
with the cumulative index $q_2$ introduced again to summarize all discrete quantum numbers describing the ground state electron and the excited nucleus. The subspace of the state vectors containing one transverse photon is associated with the projection operator 
\begin{equation}
R = \sum_r\sum_{\lambda kLM}a_{\lambda kLM}^{\dagger}|r\rangle\langle r|a_{\lambda kLM} \, ,
\end{equation}
where with $r$ we denote the quantum numbers describing the nucleus in the final state and the bound electron. Assuming the correction due to two or more photon states and due to the presence of the negative electronic continuum are negligible, we postulate the completeness relation $ P + Q_1 + Q_2 + R = \textbf{1}$. Similar decompositions involving less subspaces  for the Feshbach formalism developed for two-step NEEC and NEET  can be found in Refs.~\cite{palffy2006,Morel2004PRA}.

\subsection{\label{pertexp} Perturbation expansion of the transition operator}

The transition operator is defined as
\begin{equation}
T(z) = V + VG(z)V \,,
\end{equation}
where $V$ is the interaction Hamiltonian discussed below and  $G(z)$ is the Green operator of the system given by
\begin{equation}
G(z)= (z-H)^{-1} \,.
\end{equation}
Here, $z$ is a complex energy variable. The total Hamiltonian of the system can be written as
\begin{equation}\label{eq:totalh}
H = H_{e}+H_{n} + H_{r} +  H_{en} + H_{er} + H_{nr}\,,
\end{equation}
where with $H_{r}$ and $H_n$ we denote the  Hamiltonians describing the radiation field and   the nucleus  as discussed in Refs.~\cite{palffy2006}. The subscripts $n$, $e$ and $r$ stand for the nucleus, the electrons and the radiation field, respectively. The electronic part is given by \cite{Zimmermann}
\begin{equation}
H_{e} = \sum_{i=1}^N (c\vec{\alpha}_i\cdot \vec{p}_i+(\beta_i-1)mc^2+V_{\mathrm{nucl}}(r_i))+ \frac{1}{2} \sum_{i\ne j}\frac{1}{r_{ij}}
\, ,
\end{equation}
with $\vec\alpha$  the vector of the Dirac matrices, $\vec{p}_i$ the 
momentum of the $i$th electron and $V_{\mathrm{nucl}}$  the nuclear potential. Furthermore, 
$N$ is the number of bound electrons in the ion and $r_{ij}=|\vec{r}_i-\vec{r}_j|$. 
We adopt the Coulomb gauge for the
electron-nucleus interaction because it allows the separation of the dominant
Coulomb attraction between the electronic and the nuclear degrees of freedom \cite{palffy2006},
\begin{equation}\label{eq:coulomb}
H_{en}=\int d^3r_n\frac{\rho_n(\vec{r}_n)}{|\vec{r}_e-\vec{r}_n|}\ .
\end{equation} 
Here, $\rho_n(\vec{r}_n)$ is the nuclear charge density and
the integration is performed over the whole nuclear volume. 
As it has been shown in Ref.~\cite{palffy2006}, this Hamiltonian term describes only
the electron-nucleus interaction in transitions of electric multipolarity.
The interaction of the electron with the transverse photon field 
quantized in the volume of a sphere of radius $R$ is given by
\begin{equation}\label{eq:radhamilton}
H_{er} =-\vec{\alpha}\cdot\vec{A}=-\sum_{\lambda k LM} \left(a^{\dagger}_{\lambda kLM}\vec{\alpha}\cdot\vec{A}_{\lambda k LM}(\vec{r})+{\rm H.~c.} \right)\ ,
\end{equation}
with the vector potential of the quantized electromagnetic 
field~\cite{Ring}
\begin{equation}
\vec{A}(\vec{r})=\sum_{\lambda k LM}\left(\vec{A}_{\lambda k LM}(\vec{r})\,
a^{\dagger}_{\lambda k LM}+\vec{A}^{*}_{\lambda k LM}(\vec{r})\,a_{\lambda k LM}
\right)\ .
\end{equation} 
Here,  the two 
independent solutions of the wave equation for the $\vec{A}_{\lambda k 
LM}(\vec{r})$ are
\begin{eqnarray}\label{wave_sol}
\vec{A}_{(M)kLM}(\vec{r}) &=& \sqrt{\frac{4\pi ck}{R}}j_L(kr)
\vec{Y}^M_{LL}(	\theta,\varphi)\ ,\\
\vec{A}_{(E)kLM}(\vec{r}) &=& \frac{i}{k}\sqrt{\frac{4\pi ck}{R}}\vec{\nabla}\times\big(j_L(kr)
\vec{Y}^M_{LL}(	\theta,\varphi)\big) \ , \nonumber 
\end{eqnarray}
where the quantum number $k$ is discretized by requiring the proper 
boundary conditions at a perfectly conducting sphere of radius $R$. The 
$\vec{Y}^M_{LL}( \theta,\varphi)$ denote the vector spherical harmonics \cite{Edmonds}.

Similarly, the interaction of the nucleus with the electromagnetic field 
is given by the Hamiltonian
\begin{eqnarray}\label{eq:hnr}
H_{nr}& = &-\frac{1}{c}\sum_{\lambda k LM}  \\
&\times&
 \bigg( a^{\dagger}_{\lambda k LM}\int d^3r_n\vec{j}_n(\vec{r}_n)
\cdot\vec{A}_{\lambda k LM}(\vec{r}_n) + {\rm H.~c.} \bigg) \ ,\nonumber
\end{eqnarray}
where  $\vec{j}_n(\vec{r}_n)$ is the nuclear current.

With the help of the projector operators introduced in the previous subsection, we can separate the perturbation part in the Hamiltonian, $H=H_0+V$ with
\begin{equation}\label{eq:totalh0}
H_{0} = PHP+Q_1HQ_1+Q_2HQ_2+RHR\, ,
\end{equation}
and 
\begin{eqnarray}\label{eq:totalhint}
V &= &H-H_0\nonumber \\
&=&PVQ_1+Q_1VP+PVQ_2+Q_2VP \nonumber \\
&+& PVR+RVP+Q_1VQ_2+Q_2VQ_1 \nonumber \\
&+&  Q_1VR+RVQ_1+Q_2VR+RVQ_2\, .
\end{eqnarray}
The interaction Hamiltonian $V$ is the one responsible for the possible transitions between the four subspaces 
described in Section ~\ref{subspaces}. While $H_n$ and $H_r$ only contribute to the unperturbed Hamiltonian $H_0$, and $H_{en}$, $H_{nr}$ and $H_{er}$ only contribute to the interaction $V$, the electronic Hamiltonian $H_e$ enters both expressions. The terms $PH_eP$ and $Q_iH_eQ_i$ with $i=1,2$  describe the continuum and the bound electrons, respectively. In addition, the electron-electron interaction term in $H_e$ is also responsible for DC and Auger decay in the interaction terms $PH_eQ_1$ and $Q_1H_eP$.

The cross section of the process can be
expressed by the transition operator as follows: 
\begin{equation}\label{eq:tsigma}
\frac{d\sigma_{i \to f}}{d\Omega_k}(E) = \frac{2\pi}{F_i}
\lim_{\epsilon \to 0+}  |\langle \psi_f|T(E+i\epsilon) |\psi_i\rangle|^2 \rho_f \ ,
\end{equation}
with  $\psi_f$ and $\psi_i$ as final and initial eigenstates of $H_0$,
respectively. This cross section is differential with respect to the angle
$\Omega_k$ of the photon emitted in the process. Furthermore, $F_i$ denotes the flux of the
incoming electrons, and $\rho_f$ the density of the final photonic states. 

We use the Lippmann-Schwinger equation to write the perturbation series for $T(z)$
in powers of $V$ with the Green function $G_0(z)$ of the unperturbed Hamiltonian $H_0$:
\begin{equation}
T(z) = V + VG_0(z)V + VG_0(z)VG_0(z)V + \dots \,.
\end{equation}
Since the initial state of the DC process is by definition an eigenstate of
$P$, and the final state after the $\gamma$ decay  is an eigenstate of $R$, we consider the projection
$RTP$ of the transition operator:
\begin{eqnarray}\label{eq:rtp}
RTP = RVP &+& RVG_0VP + RVG_0VG_0VP 
\nonumber \\
&+&RVG_0VG_0VG_0VP + \dots  
\end{eqnarray}
Here and in the following we omit the argument $z$. The first term in Eq.\ (\ref{eq:rtp}) accounts for RR and does not contribute to the  three-step process under consideration. The second term $RVG_0VP= RH_{er}Q_1G_0Q_1H_{ee}P+RH_{nr}Q_2G_0Q_2H_{en}P$ accounts for the first order of two possible processes: DC followed by x-ray emission, which is DR and proceeds without any participation of the nucleus, and NEEC followed by radiative decay of the nucleus for the case of electric multipole transitions \cite{palffy2006}. Inserting the spectral resolution (\ref{Q2spectral}) of $Q_2$ in the second order in $V$ for the NEEC term we arrive to
\begin{equation}\label{eq:rtplowest}
\langle \psi_f |RTP| \psi_i \rangle =
\sum_{q_2}
\frac{\langle \psi_f | H_{nr} |q_2 \rangle\langle q_2 | H_{en}
|  \psi_i\rangle }{z-E_q^0} \ .
\end{equation}
The energy $E_q^0$ denotes the unperturbed eigenvalue of the state $|q\rangle$.

The third order term $RVG_0VG_0VP$ can be  decomposed as:
\begin{eqnarray*}
 RV\textbf{1}G_0\textbf{1}V\textbf{1}G_0\textbf{1}VP & = & RVPG_0PVQ_1G_0Q_1VP \\
&+& RVPG_0PVQ_2G_0Q_2VP \\ 
&+& RVPG_0PVRG_0RVP \\
&+& RVQ_1G_0Q_1VQ_2G_0Q_2VP \\
&+& RVQ_1G_0Q_1VRG_0RVP \\
&+& RVQ_2G_0Q_2VQ_1G_0Q_1VP \\
&+& RVQ_2G_0Q_2VRG_0RVP\, .
\end{eqnarray*} 
The sixth term  
\begin{eqnarray}
&& RVQ_2G_0Q_2VQ_1G_0Q_1VP = \nonumber \\
&& \underbrace{RH_{nr}Q_2}_{\gamma}G_0\underbrace{Q_2H_{en}Q_1}_{NEET\,  (electric)}G_0\underbrace{Q_1H_{ee}P}_{DC}
\end{eqnarray}
describes DC followed by NEET of electric transitions and contributes to the cross section of the considered process. Among the decomposition terms in the sum above we can also find the one responsible for NEEC for magnetic nuclear transitions \cite{palffy2006}. For DC followed by NEET, the term responsible for transitions of magnetic multipolarity will only emerge from the decomposition of the fourth order term in the perturbation expansion and has the form 
\begin{equation}
\underbrace{RH_{nr}Q_2}_{\gamma}G_0\underbrace{Q_2H_{nr}RG_0RH_{er}Q_1}_{NEET\, (magnetic)}G_0\underbrace{Q_1H_{ee}P}_{DC}\, .
\end{equation}
We have used here the result 
\begin{equation}
QH_{nr}RG_0RH_{er}P=QH_{\mu}P\, ,
\end{equation}
with the Hamilton operator
\begin{equation}
H_{\mu} = - \frac{1}{c} \vec{\alpha} \int d^3r_n \frac{\vec{j}_n(\vec{r}_n)}{|\vec{r}-\vec{r}_n|}
=-\vec{\alpha}\cdot\vec{A}(\vec{r})
\label{hmagn}
\end{equation}
obtained in Ref.~\cite{palffy2006} which accounts for a bound electron that decays by
exchanging a virtual transverse photon with the nucleus.

The fourth order decomposition  in the perturbation expansion also contains the exchange of a virtual photon between a bound and a continuum photon, i.e., the Breit interaction in the term
\begin{equation}
\underbrace{RH_{nr}Q_2}_{\gamma}G_0 \underbrace{Q_2H_{en}Q_1}_{NEET\, (electric)} G_0 \underbrace{Q_1H_{er}RG_0 RH_{er}P}_{DC \,(Breit)}\, .
\end{equation}
Together with the Coulomb electron-electron interaction term in the interaction Hamiltonian, the Breit term contributes to DC and  Auger decay. 
Using results well known in the theory of DR \cite{Zimmerer,Zimmermann} we can write the Hamiltonian responsible for DC and Auger decay as the sum of the Coulomb and the Breit contributions
\begin{eqnarray}
H_{ee} &=&\frac{1}{2} \sum_{i\ne j}\frac{1}{r_{ij}}- \sum_{i\ne j}\vec{\alpha}_i\cdot \vec{\alpha}_j\frac{\cos (\omega r_{ij})}{r_{ij}}
\nonumber \\
&+&(\vec{\alpha}_i\cdot\vec{\nabla}_i)(\vec{\alpha}_j\cdot\vec{\nabla}_j)\frac{\cos (\omega r_{ij})-1}{\omega^2 r_{ij}}
\ .
\end{eqnarray}
where $\omega$ is the frequency of the exchanged virtual photon.

Following the procedure described in Ref.~\cite{palffy2006} and adopting the so-called isolated resonance approximation we can continue with the next terms in the expansion up to infinite order. In Ref.~\cite{palffy2006}, the infinite perturbation expansion was shown to introduce energy and width corrections into the energy denominator of the lowest order amplitude. The sum over all orders could be performed with the help of a geometric progression. In this case, since we have two projector operators  $Q_1$ and $Q_2$ corresponding to two types of intermediate states, the infinite perturbation expansion can be reduced to summing two geometrical progressions to obtain 
\begin{eqnarray}
&&\langle\psi_f|RT(z)P|\psi_i\rangle = \\
&&\frac{\sum_{d_1d_2}\langle\psi_f|H_{nr}|\psi_{d_2}\rangle\langle\psi_{d_2}|H_{N}|\psi_{d_1}\rangle\langle\psi_{d_1}|H_{ee}|\psi_i\rangle}{\left(z - E_{d_2} + \frac{i}{2}\Gamma_{d_2}\right)\left(z - E_{d_1} + \frac{i}{2}\Gamma_{d_1}\right)}\, , \nonumber 
\end{eqnarray}
with the notation $H_N=H_{en}+H_{\mu}$.
The energies $E_{d_1}$ and $E_{d_2}$ of the electronic and nuclear transitions, respectively, are including the electron-electron interaction and the corresponding  Breit corrections and all the electronic and nuclear self-energy contributions, as well as the nuclear polarization contributions for the studied nuclear transition, $E_{d_i}=E^0_{d_i}+\Delta E_{d_i}^{\rm{NP}} +\Delta E_{d_i}^{\rm{Br}}+ \Delta E_{d_i}^{\rm{SE}} + \Delta E_{d_i}^{\rm{NSE}}$ \cite{palffy2006}. Correspondingly, the widths $\Gamma_{d_2}$ (nuclear) and $\Gamma_{d_1}$ (atomic) include electronic radiative and Auger widths for state $d_1$ as well as nuclear radiative and IC widths for state $d_2$. 

Finally, the total cross section for our three-step process, after summing over all final states and averaging over all initial states will has form
\begin{widetext}
\begin{equation}
\sigma(E)=\frac{2\pi}{F_i}\frac{1}{2J_i+1}\sum_{M_i}\frac{1}{4\pi}\int d\Omega_{\vec{p}} \frac{1}{2} \sum_{m_s}\frac{1}{2I_g+1}\sum_{M_g,M_2,M_f}\sum_{L,M,\lambda} 
 \left| \sum_{d_1,d_2} \frac{\langle\psi_f|H_{nr}|\psi_{d_2}\rangle\langle\psi_{d_2}|H_{N}|\psi_{d_1}\rangle\langle\psi_{d_1}|H_{ee}|\psi_i\rangle}{\left(E-E_{d_1}+\frac{i}{2}\Gamma_{d_1}\right)\left(E-E_{d_2}+\frac{i}{2}\Gamma_{d_2}\right)}\right|^2 \rho_f \, .
\end{equation}
\end{widetext}
%

\subsection{Resonance strength for the three-step process }

The resonance strength (also known as  integrated cross section) is  obtained by integrating the total cross section over the whole energy spectrum. 
A simplification of its expression can be obtain by studying the relation between the two widths $\Gamma_{d_1}$ and $\Gamma_{d_2}$ that appear in the total cross section denominator. For the considered case, the final electronic state is the ground state. Since the nuclear excited state width is much smaller than the electronic excited state width,  $\Gamma_{d_1}$ is approximately given by the electronic width of the initial state and $ \Gamma_{d_2}$ is the nuclear width, such that $\Gamma_{d_1}\gg \Gamma_{d_2}$. 
Neglecting the variation of the free electron momentum and matrix elements on the energy interval of interest determined by the two Lorentzian profiles and dominated by the very narrow nuclear width, the integration resumes to
\begin{eqnarray}
\int \frac{dE}{\left((E-E_{d_1})^2+\frac{\Gamma^2_{d_1}}{4}\right)\left((E-E_{d_2})^2+\frac{\Gamma^2_{d_2}}{4}\right)}=
\nonumber \\
\frac{2\pi}{\Gamma_{d_2}\left((E_{d_1} - E_{d_2})^2 + \frac{\Gamma^2_{d_1}}{4}\right)}\, .
\end{eqnarray}
 The product between the flux of the incoming electrons $F_i$ and the density 
of the initial electronic states $\rho_i$ does not depend on the normalization of the continuum wave functions \cite{Zimmerer},
\begin{equation}
F_i\rho_i=\frac{p^2}{(2\pi)^3}\, .
\end{equation}
The resonance strength of the three-step process can be then written as a product of
three terms,
\begin{equation}
\int \sigma(E) dE=\frac{\pi^2}{p^2} \frac{2J_1+1}{2J_i+1}\frac{A_{Au} A_{\gamma}\mathcal{P}_{neet}}
{\Gamma_{d_2}}\, ,
\end{equation}
where we have introduced the Auger decay rate
\begin{eqnarray}
A_{Au}&=&\frac{2\pi}{2J_1+1}\sum_{M_1,m_s,M_i} 
\nonumber \\
&\times &\int d\Omega_{\vec{p}} |\langle \alpha_1 J_1 M_1|H_{ee}| \alpha_i J_i M_i\vec{p} m_s\rangle|^2 \rho_i\, ,
\end{eqnarray}
the NEET probability 
\begin{eqnarray}
\label{pneet}
\mathcal{P}_{neet}&=&\frac{1}{(2I_g+1)(2J_1+1)}\sum_{M_1,M_2,M_e,M_g}
 \\
&\times& \frac{|\langle  I_e M_e,\alpha_2 J_2 M_2|H_{N}| I_g M_g, \alpha_1 J_1 M_1\rangle|^2
 }{(E_{d_1}-E_{d_2})^2+\frac{\Gamma_{d_1}^2}{4}}\, ,\nonumber
\end{eqnarray}
and finally the radiative decay rate of the nuclear excited state,
\begin{eqnarray}
A_{\gamma}&=&\frac{2\pi}{2I_e+1}\sum_{M_e,M_f,\lambda k LM}
\nonumber  \\
&\times&  |\langle  I_g M_f,\lambda k LM |H_{nr}|  I_e M_e\rangle|^2 \rho_f\, .
\end{eqnarray}
Our formalism reproduces the results presented in Ref.~\cite{Tkalya1992,Ho1993} for the NEET probability for the particular case when the second intermediate state $d_2$ is the  electronic ground state. 
Determining the total cross section of the studied process requires the
calculation of the transition rates $A_{Au}$ and $A_{\gamma}$, the NEET probability $\mathcal{P}_{neet}$  and the initial and final
state energies. In the actual calculations we neglect the additional nuclear polarization corrections
$\Delta E_d^{\rm{NP}}$  to the bound electron energies \cite{MPSoff}, since their value is smaller than the nuclear transition energy uncertainties and not known with good accuracy.


\section{\label{rates} The NEET probability for electric and magnetic transitions}
The calculation of the NEET probability requires knowledge of the atomic levels and widths 
and the evaluation of the interaction matrix element in the numerator of Eq.~(\ref{pneet}).
In the following we sketch the calculation of the interaction matrix element
$\langle  I_e M_e,\alpha_2 J_2 M_2|H_{N}| I_g M_g, \alpha_1 J_1 M_1\rangle$
following the approach used for the NEEC rates \cite{palffy2006}. 

In order to calculate the NEET probability we have considered the matrix element of the
electric and magnetic interactions between the electron and the nucleus. We
write the wave function of the system as the product wave function of the
electronic and nuclear states.
Unlike for NEEC, now we have in both initial and final state only bound electrons. The
electronic multiconfiguration wavefunctions $| \alpha_1 J_1 M_1\rangle$ are obtained by Slater determinants constructed with
relativistic orbitals $|n\kappa m\rangle$ in the framework of the multiconfiguration Dirac-Fock method (MCDF). The relativistic orbitals are eigenfunctions of the angular-momentum operators $\hat{j}^2$ and $\hat{j}_z$ and of the parity operator for each electron. The notation $n$ accounts for the principal quantum number, $\kappa$ is the relativistic angular quantum number $\kappa=\pm (j+1/2)$ for $l=j\pm 1/2$ such that $j=|\kappa|-1/2$, and $m$ is the projection of the angular momentum operator $j$ on the $z$ axis. 

The electron-nucleus interaction Hamiltonian describing electric multipolarity transitions can be written using the multipole expansion as
\begin{equation}
H_{en} = \sum_{L = 0}^{\infty}\sum_{M = -L}^{L}Y_{LM}^*(\Omega_e)\int d^3r_n\frac{r_<^L}{r_>^{L + 1}}Y_{LM}(\Omega_n)\rho_n(\vec{r}_n)\, .
\end{equation}
The radius $r_e(r_n)$ denotes the electronic (nuclear) radial coordinate and $\Omega_e(\Omega_n)$ stands for the corresponding solid angle, while the $Y_{lm_l}(\Omega_e)$
denote the spherical harmonics. We  make the simplifying assumption that the electron does not enter the nucleus, such that $r_e > r_n$. The Hamiltonian can then be written as 
\begin{equation}
H_{en} = \sum_{L = 0}^{\infty}\sum_{M = -L}^{L}\frac{4\pi}{2L + 1}Y_{LM}^*(\Omega_e)\frac{1}{r_e^{L + 1}}Q_{LM}
\end{equation}
with the help of electric multipole moments $Q_{LM} = \int d^3r_nr_n^LY_{LM}(\Omega_n)\rho_n(\vec{r}_n)$. The nuclear part of the matrix element can then be structured in the reduced electric transition probability
\begin{equation}
B\!\uparrow \!(EL, I_g \rightarrow I_e) = \frac{1}{2I_g + 1}|\langle I_e||Q_L||I_g\rangle|^2\, .
\end{equation}
Using the single active-electron approximation and following the calculation in Ref.~\cite{palffy2006}, we obtain for the NEET probability for electric transitions
\begin{eqnarray}
\label{pneetE1}
P_{neet}^{(E)}& =& \frac{4\pi}{(2L + 1)^2}\frac{(2J_{2} + 1)}{(2J_{1} + 1)}B\!\uparrow \!(EL, I_g \rightarrow I_e) \\
&\times&C\left(J_{2}\ L\ J_{1};\frac{1}{2}\ 0\ \frac{1}{2}\right)^2 \frac{|R^E_{L,J_{1},J_{2}}|^2}{(E_{d_1} - E_{d_2})^2 + \frac{\Gamma_{d_1}^2}{4}}\, .\nonumber
\end{eqnarray}
Here, $C\left(J_{2}\ L\ J_{1};\frac{1}{2}\ 0\ \frac{1}{2}\right)$ stands for the vector coupling coefficients and the radial integral $R^{(E)}_{L,J_{1},J_{2}}$ is given by
\begin{eqnarray}
\label{radint}
R^{(E)}_{L,J_{1},J_{2}} &=& \int_{0}^{\infty}drr^{-L + 1}\left[f_{n_{2}\kappa_{2}}(r)f_{n_{1}\kappa_{1}}(r) \right. 
\nonumber \\
&+& \left. g_{n_{2}\kappa_{2}}(r)g_{n_{1}\kappa_{1}}(r)\right]\, ,
\end{eqnarray}
where $g_{n\kappa}(r)$ and $f_{n\kappa}(r)$ are the large and small components of the bound Dirac radial wave functions describing the orbitals of the active electron
\begin{equation}
\psi_{n\kappa m}(\vec{r}) = \left(\begin{array}{r}g_{n\kappa}(r)\Omega_{\kappa}^{m}(\Omega_e) \nonumber \\ if_{n\kappa}(r)\Omega_{-\kappa}^{m}(\Omega_e)\end{array}\right)\, ,
\end{equation}
and the indices $1$ and $2$ denote the two states  $\psi_{d_1}$ and $\psi_{d_2}$, respectively.

For magnetic multipole transitions, 
the magnetic Hamiltonian in Eq.~(\ref{hmagn}) can be written using the multipole
expansion as
\begin{eqnarray}
&&H_{\mu}=-\vec{\alpha}\cdot\vec{A}=-\frac{1}{c}\sum_{LM}\frac{4\pi}{2L+1}\vec{\alpha}\cdot
\vec{Y}^M_{LL}(\Omega_e) 
\nonumber \\
&&\int
d^3r_n\frac{r_<^L}{r_>^{L+1}}\vec{j}_n(\vec{r}_n)\cdot\vec{Y}^{M*}_{LL}(\Omega_n)\ .
\end{eqnarray}
Using also in this case the approximation that the electron does not enter the nucleus, 
we obtain
\begin{equation}
 H_{\mu}=-i\sum_{LM}\frac{4\pi}{2L+1}\sqrt{\frac{L+1}{L}}
r_e^{-(L+1)}M_{LM}\vec{\alpha}\cdot\vec{Y}^{M*}_{LL}(\Omega_e)\, ,
\label{hmagnmult}
\end{equation}
where  the magnetic multipole operator is given by \cite{Schwartz} 
\begin{equation}
M_{LM}=-\frac{i}{c}\sqrt{\frac{L}{L+1}}\int d^3r
r^L\vec{Y}^{M}_{LL}(\Omega_n)\cdot\vec{j}_n(\vec{r}_n)\, .
\end{equation}
Similar to the case of the electric transitions, 
all the nuclear information can be contained in the reduced magnetic transition
probability
\begin{equation}
B\!\uparrow\!(ML,I_g\to I_e)=\frac{1}{2I_g+1}|\langle  I_e\|M_L\|I_g\rangle|^2\ ,
\end{equation} 
whose value can be taken from experimental data or from theoretical calculations considering
different nuclear models. After some angular momentum algebra 
we obtain for $\mathcal{P}_{neet}$
\begin{eqnarray}
P_{neet}^{(M)} &=& \frac{4\pi}{L^2(2L + 1)^2} \frac{(2J_2+1)}{(2J_{1} + 1)}B\!\!\uparrow\!\!(ML, I_i\rightarrow I_d)
\nonumber \\
&\times& (\kappa_1 + \kappa_2)^2 C\left(J_{2}\ L\ J_{1};\frac{1}{2}\ 0\ \frac{1}{2}\right)^2  \left|R_{L,J_1,J_2}^{(M)}\right|^2  \nonumber \\
&\times&
\frac{1}{(E_{d_1} - E_{d_2})^2 + \frac{\Gamma_{d_1}^2}{4}}\, .
\end{eqnarray}
The electronic radial integral is in this case
\begin{eqnarray}
R_{L,J_1,J_2}^{(M)}&=&\int_0^{\infty}drr^{-L + 1}\left[g_{n_2\kappa_2}(r_e)f_{n_2\kappa_1}(r)  \right.
\nonumber \\
&+& \left. f_{n_2\kappa_2}(r)g_{n_1\kappa_1}(r)\right]\, .
\end{eqnarray} 
Given the different parity of the electric and
magnetic multipole moments, a transition of a given multipolarity $L$ is either
electric or magnetic. We consider in the following only the cases of transitions with a certain
value of $L$ which do not present mixing ratios between electric and
magnetic multipoles of different multipolarities.

\section{Interference terms \label{interf}}
\begin{figure}[h]
  \centering
  \includegraphics*[width=8.7cm]
             {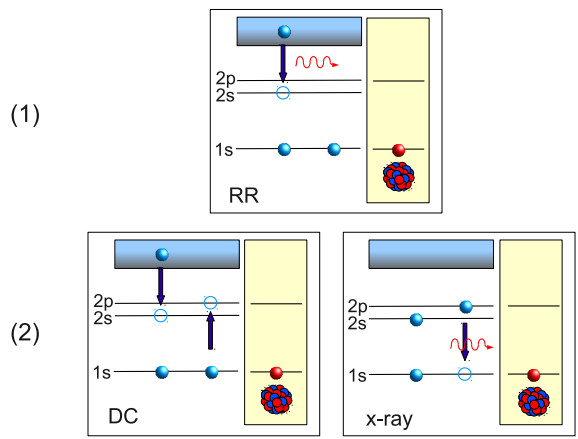}
  \caption{  Schematic picture of the two  interference pathways involving purely atomic processes: the direct process of RR (1) and DC + x-ray decay (2). }
  \label{figure2}
\end{figure}

\begin{figure}[h]
  \centering
  \includegraphics*[width=8.7cm]
             {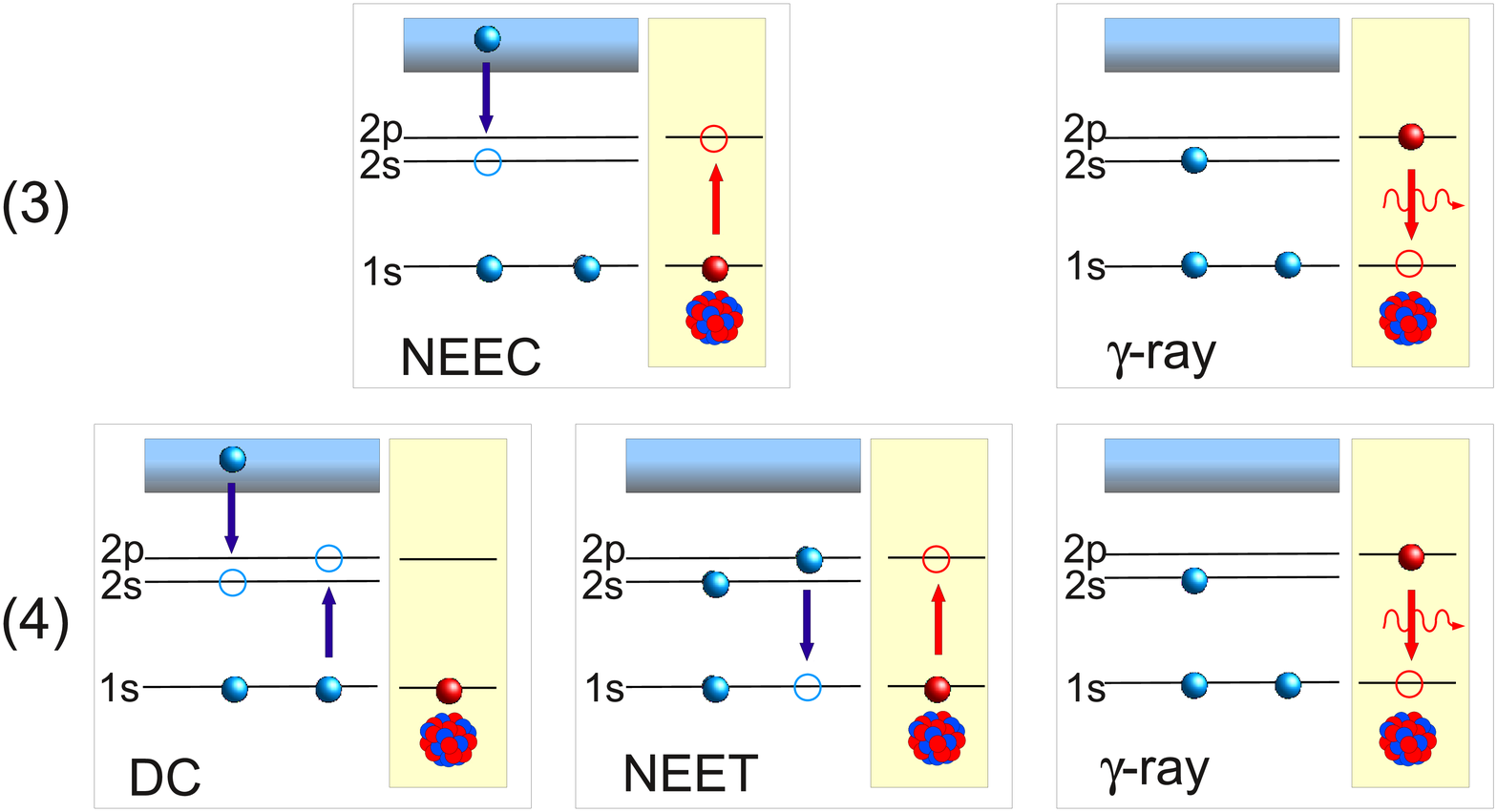}
  \caption{ Schematic picture of the two interference pathways that couple the atomic and nuclear degrees of freedom: the two step process NEEC + $\gamma$ decay (3) and  the three-step process of DC + NEET + $\gamma$ decay (4).}
  \label{figure3}
\end{figure}

As already mentioned in the introduction, our three step process consisting of DR, NEET and
$\gamma$-ray emission has a number of competing processes in each step that open a number of interference
channels. Interference can occur whenever the initial and final states of two competing processes are the same. 
For exemplification, we will discuss in the following a suitable case in which interference between all electron recombination 
and nuclear excitation mechanisms may occur. This is the case for DC occurring  in He-like HCI---in  our numerical case we have considered capture of a free electron into He-like
$^{237}\mathrm{Np}$ which has a suitable nuclear transition for NEET. The four interference pathways are presented in Figs.~\ref{figure2} and \ref{figure3}: (1) RR, (2) DC + x-ray decay, i.e., DR,  (3) NEEC + $\gamma$ and (4) DC + NEET + $\gamma$. Here for brevity we have replaced ``followed'' by the sign ``$+$''.


In the common initial state  $\psi_i$ we have a He-like ion in the electronic ground state $1s^2$ with its nucleus in the ground state, and a free electron of  energy $E_{\vec{p}}$ and angular momentum quantum numbers of the partial wave $|\varepsilon\kappa jm\rangle$. The final state consists of the  ground state Li-like ion, ground state nucleus, and an $E1$ photon of energy matching the nuclear and atomic transitions. As long as the state $1s2s2p_{3/2}$ decays directly to the ground state (consistent with the $1s2s2p_{3/2}\rightarrow 1s^22p_{3/2}$  transition rate  being several orders of magnitude smaller than the $1s2s2p_{3/2}\rightarrow 1s^22s$ one), the four pathways cannot be discerned and thus quantum interference occurs.

The Hamiltonian terms that describe the four interference pathways can be found in the perturbation expansion of the transition operator presented in Section~\ref{pertexp}. Except for the direct process of RR, which is described by the term $RVP$, all the other three combinations include resonant processes that can only be accounted for in the perturbation expansion by summing the corresponding terms up to infinite order. This procedure, originally developed for the calculation of DR cross sections, has been also adapted for NEEC \cite{palffy2006,interference2007} and for NEET in this work. Thus, we can write  the transition operator of the four interference pathways that connect the initial and the final states $\psi_i$ and $\psi_f$ as
\begin{eqnarray}
\label{interfTME}
&&\langle \psi_f |RT(z)P|\psi_i \rangle =
\langle\psi_f |RH_{er}P|\psi_i\rangle \nonumber \\
&&+\sum_{d'_1d'_2}\frac{\langle\psi_f|H_{nr}|\psi_{d'_2}\rangle\langle\psi_{d'_2}|H_{N}|\psi_{d'_1}\rangle\langle\psi_{d'_1}|H_{ee}|\psi_i\rangle}{(z - E_{d'_2} + \frac{i}{2}\Gamma_{d'_2})(z - E_{d'_1} + \frac{i}{2}\Gamma_{d'_1})}\nonumber \\
&&+\sum_{d_2}\frac{\langle \psi_f | H_{nr} |\psi_{d_2} \rangle
\langle \psi_{d_2}| H_{N} |\psi_i \rangle}
{z-E_{d_2} + \frac{i}{2} \Gamma_{d_2}} \nonumber \\
&&+\sum_{d_1}\frac{\langle \psi_f | H_{er} |\psi_{d_1} \rangle
\langle \psi_{d_1}| H_{ee} |\psi_i \rangle}
{z-E_{d_1} + \frac{i}{2} \Gamma_{d_1}} 
\, .
\end{eqnarray}
By introducing the expression above in the  cross section formula in Eq.~(\ref{eq:tsigma}), we obtain a sum of ten terms, four of them corresponding to the cross sections of the four considered processes pathways and the other six to  the interference between them.
In the following we label the simple terms with $\sigma_i$ and the interference terms between pathways $i$ and $j$ as $\sigma_{ij}$, with $i,j=1,2,3,4$.
 We are now interested in the three interference terms that involve the process under consideration, labeled  as channel (4). Note that two of the other three interference terms, $\sigma_{12}$ and $\sigma_{13}$, have been already investigated in Refs.~\cite{interference2007,Zimmermann}. 

In the following we will address the test case of $^{237}\mathrm{Np}$ and sketch the calculation of the three interference terms. The NEET nuclear transition  is occurring between the ground state and the excited state at 102.959 keV. As it will be shown in the numerical results section, this energy is best matched by the $E1$ atomic transitions between the $1s2s2p_{3/2}$ state and the ground state $1s^22s$ of the Li-like ion. The active electron is undergoing a transition from the $2p_{3/2}$ to the $1s$ orbital. The decay of the $2s$ electron to the $K$-shell occurs much slower and thus  justifies the considered one active electron scenario. The DC occurs in the ground state He-like ion, such that the free electron is captured in the $2s$ orbital with the simultaneous excitation of one of the $1s^2$ electrons to the $2p_{3/2}$ state.
Since this is an $E1$ transition, both RR and DR processes have large cross sections, such that the corresponding interference term in the cross section is expected to be relevant for the total cross section. The numerical study in the next section  shows to what extend this is  the case.

\subsection{RR interference term}
In order for interference between RR and pathway (4) to occur, the continuum electron that recombines via RR or DC has to have the same initial and final energies and quantum numbers. In both cases, the recombination occurs into the $2s$ orbital. In RR the final electronic bound ground state is reached (the nucleus remains in the ground state all the  time), while for the three-step process (4) the recombination into the $2s$ orbital leads to the excited state $1s2s2p_{3/2}$ that decays via NEET. 

By introducing the expression of the transition operator matrix element in Eq.~(\ref{interfTME}) in the expression of the total cross section, we obtain the interference cross section term between pathways (1) and (4),
\begin{widetext}
\begin{eqnarray}
&&\sigma_{14}(E)=\frac{(2\pi)^4}{p^2}\rho_i\rho_f\sum_{M,M_f,M_2}\frac{1}{4\pi}\int d\Omega_{\vec{p}}\frac{1}{2}\sum_{m_s}\frac{1}{2J_i+1}\sum_{M_i}\frac{1}{2I_g+1}\sum_{M_g} \Bigg(\langle \alpha_2 J_2 M_2,(E)kLM|H_{er}|\alpha_i J_i M_i \vec{p} m_s\rangle^*  \\
&&\times \left. \sum_{M'_e}\sum_{\alpha'_1J'_1 M'_1}\frac{\langle I_g M_f,(E)kLM|H_{nr}|I_e M'_e\rangle \langle I_e M'_e,\alpha_2 J_2 M_2|H_{en}|I_g M_g,\alpha'_1 J'_1 M'_1\rangle \langle \alpha'_1J'_1 M'_1|H_{ee}|\alpha_i J_i M_i \vec{p} m_s\rangle}
{\left(E-E_{d'_1}+\frac{i}{2}\Gamma_{d'_1}\right)\left(E-E_{d_2}+\frac{i}{2}\Gamma_{d_2}\right)} + \mathrm{H.~c.}\right)\, .\nonumber 
\end{eqnarray}
\end{widetext}
Here we have assumed that NEET can occur only via a specific nuclear level denoted by the total angular momentum  $I_e$ (whereas its projection is not fixed and thus must be summed over). Furthermore, for the sake of brevity, the non-participating  electronic or nuclear wavefunctions in the matrix elements of $\gamma$ decay or DC, respectively, were omitted in the notation. 
The partial wave expansion of the continuum wave function reads
\begin{eqnarray}
|\vec{p} m_s\rangle&=&\sum_{\kappa m m_l}i^l e^{i\Delta_{\kappa}} Y_{l m_l}^*(\Omega_p) \nonumber \\
&\times &
C\left(l\ \frac{1}{2}\ j;m_l\ m_s \ m\right)| \varepsilon\kappa j m\rangle\, ,
\end{eqnarray}
where $\varepsilon$ is the energy of the continuum electron measured 
from the ionization threshold, $\varepsilon=\sqrt{p^2c^2+c^4}-c^2$. The 
orbital angular momentum of the partial wave is denoted by $l$ and the 
corresponding magnetic quantum number by $m_l$, while the partial wave 
phases $\Delta_{\kappa}$ are chosen so that the continuum wave function 
fulfills the boundary conditions of an incoming plane wave and an 
outgoing spherical wave. The total angular momentum quantum number of 
the partial wave is $j=|\kappa|-\frac{1}{2}$. 
The  expression of the RR matrix element (see, for instance, Refs.~\cite{Eichler2000,interference2007}) is given by
\begin{eqnarray}\label{eler}
&&\langle \alpha_2 J_2 M_2, (E) kLM|H_{er}|\varepsilon \kappa j m  \rangle \nonumber \\
&=&i(-1)^{j-L+\frac{1}{2}}\sqrt{\frac{4\pi ck}{R}}C(j\ L\ J_2;m\ M\ m_d)
\sqrt{\frac{2j+1}{4\pi}}
\nonumber \\
&\times & 
\left(\begin{array}{ccc}J_2 & j& L\\\frac{1}{2}&-\frac{1}{2}&0\end{array}\right)
R^{RR}_{J_2,j,L}\, .
\end{eqnarray}
The Wigner $3j$-symbol used in the expression above is related to the Clebsch-Gordan coefficients via
\begin{eqnarray}
C(j_1\ j_2\ j;m_1\ m_2\ m)&=&(-1)^{m+j_1-j_2}\sqrt{2j+1}
\nonumber \\
&\times&
\left(\begin{array}{ccc} j_1&j_2&j\\m_1&m_2&-m\end{array}\right)\, .
\end{eqnarray}
The expression  $R^{RR}_{J_2,j,L}$ involving the radial integrals is in this case given by
\begin{eqnarray}
\label{rrme}
&&R^{RR}_{J_2,j,L}=
 \bigg[\sqrt{\frac{L+1}{L(2L+1)}}(LI^-_{L-1}-(\kappa_d-\kappa)I^+_{L-1})
 \\
&&
  +  \sqrt{\frac{L}{(L+1)(2L+1)}}
((L+1)I^-_{L+1}+(\kappa_d-\kappa)I^+_{L+1})\bigg]\ , \nonumber
\end{eqnarray}
with the radial integrals
\begin{equation}\label{eq:iint}
I^{\pm}_L=\int_0^{\infty}drr^2j_L(kr)\left(
g_{n_2 \kappa_2}(r)f_{\varepsilon\kappa}(r)
\pm g_{\varepsilon\kappa}(r)f_{n_2 \kappa_2}(r)\right)\,,
\end{equation}
containing the spherical Bessel functions $j_L(kr)$ and the  large and small radial components of
the relativistic continuum electron partial wave function $g_{\varepsilon\kappa}(r)$ and $f_{\varepsilon\kappa}(r)$, respectively.

For the matrix element of the interaction Hamiltonian 
between the nucleus and the radiation field $H_{nr}$, we make the assumption that the wavelength of the radiation is 
large compared to the nuclear radius, $kR_0\ll 1$, so that the Bessel 
functions that appear in the expression of the field vector potential (\ref{wave_sol}) can be approximated in the first order in $kr$ as
\begin{equation}
j_L(kr)\simeq \frac{(kr)^L}{(2L+1)!!}\, .
\label{lwl}
\end{equation}
In this case the electric solution of the wave equation can be written as
\begin{eqnarray}
\vec{A}_{(E)kLM}(\vec{r})&=& -\sqrt{\frac{4\pi ck}
{R}}\frac{\sqrt{(L+1)(2L+1)}}{(2L+1)!!}
\nonumber \\
&\times & (kr)^{L-1}\vec{Y}^M_{LL-1}(\theta,\varphi)
\, .
\end{eqnarray}
With the use of the continuity equation for the nuclear current 
$\vec{j}_n$ we obtain for the matrix element
\begin{eqnarray}\label{elnr}
&&\langle I_g M_{f},(E)kLM |H_{nr}|I_e M_{e}\rangle
\nonumber \\
&&=(-1)^{I_e-M_{e}+1}\sqrt{\frac{4\pi ck}{R}}
C\left(I_g\ I_e\ L; M_{g}\ -M_{e}\ M\right)  \nonumber \\
&&\times \frac{\sqrt{L+1}}{\sqrt{L(2L+1)}}\frac{ik^L}{(2L+1)!!}
\langle I_g  \|Q_L\| I_e \rangle \,.
\end{eqnarray}
Finally, one can relate in a simple manner the DC matrix element to the Auger rate for the case under consideration, making the observation that 
since the total angular momentum of the bound Auger final state $1s^2$ is 0,  there is only one corresponding partial wave for the recombining continuum electron with $j=1/2$ ($j=3/2$) and $\kappa=1$ $(\kappa=-2)$ for the $J_1=1/2$ $(J_1=3/2)$ intermediate state. 
We then have
\begin{eqnarray}
\label{augME}
&&\langle \alpha_1 J_1 M_1|H_{ee}|[\alpha_i J_i M_i j m]_{J'_1M'_1}\rangle =
\nonumber \\
&& \langle \alpha_1 J_1 \|H_{ee}\|[\alpha_i J_i  j ]_{J'_1}\rangle \delta_{J_1J'_1}
\delta_{M_1M'_1}\, ,
\end{eqnarray}
and the Auger rate is given by
\begin{equation}
A_{Au}=2\pi
 |\langle \alpha_1 J_1 \|H_{ee}\| [\alpha_i J_i  j ]_{J_1}\rangle|^2 \rho_i\, .
\end{equation}
Combining the expressions above, we obtain for the interference  cross section the expression
\begin{eqnarray}
\label{rrinterf}
\sigma_{14}(E)&=&\frac{16\pi^3}{p^2}\rho_iB\!\uparrow\!(E1,I_g\to I_e)(2J_2+1)(-1)^{3I_e+I_g} \nonumber \\
&\times& \frac{k^{L+1}}{(2L+1)!!}\sqrt{\frac{L+1}{L((2L+1)^3}}\left(\begin{array}{ccc}J_2 & j& L\\\frac{1}{2}&-\frac{1}{2}&0\end{array}\right)^2
\nonumber \\
&\times& R_{L,J_1,J_2}^{(E)}
R^{RR}_{J_2,j,L}
\langle \alpha_1 J_1 \|H_{ee}\| [\alpha_i J_i  j ]_{J_1}\rangle
 \\
&\times& \mathrm{Re}\left(\frac{1}{\left(E-E_{d_1}+\frac{i}{2}\Gamma_{d_1}\right)\left(E-E_{d_2}+\frac{i}{2}\Gamma_{d_2}\right)}\right)\, .\nonumber
\end{eqnarray}
%
\subsection{DR interference term}
DC is the first step for both pathways (2) and (4). By DC, the doubly-excited electronic state $1s2s2p_{3/2}$ is reached, that can subsequently decay either by NEET or by x-ray emission. NEET may occur only when the $2p$ electron decays to the $1s$ hole, while the $2s$ electron acts as spectator. In principle,  the $2s$ electron may also decay by x-ray emission to the $K$-shell and contribute to DR, but this process is much slower. We therefore consider in the following the interference between NEET followed by $\gamma$ decay and $E1$ x-ray emission of the doubly-excited electronic state  following DC. We can write the interference term in the cross section as
\begin{widetext}
\begin{eqnarray}
&&\sigma_{24}(E)=\frac{(2\pi)^4}{p^2}\rho_i\rho_f\sum_{M,M_f,M_2}\frac{1}{4\pi}\int d\Omega_{\vec{p}}\frac{1}{2}\sum_{m_s}\frac{1}{2J_i+1}\sum_{M_i}\frac{1}{2I_g+1}\sum_{M_g}  \\
&&\times 
\left(\frac{ \sum_{\alpha_1J_1M_1} \langle \alpha_2 J_2 M_2,(E)kLM |H_{er}|\alpha_1 J_1 M_1\rangle^*\langle \alpha_1 J_1 M_1|H_{ee}| \alpha_i J_i M_i \vec{p} m_s\rangle^*}{E-E_{d_1}-\frac{i}{2}\Gamma_{d_1}}
\right. \nonumber \\
&&\times  \left.
\sum_{M'_e}\sum_{\alpha'_1J'_1 M'_1}\frac{\langle I_g M_f,(E)kLM |H_{nr}|I_e M'_e\rangle \langle I_e M'_e,\alpha_2 J_2 M_2|H_{en}|I_g M_g, \alpha'_1 J'_1 M'_1\rangle \langle \alpha'_1 J'_1 M'_1|H_{ee}|\alpha_i J_i M_i \vec{p} m_s\rangle}
{\left(E - E_{d'_1}+\frac{i}{2}\Gamma_{d'_1}\right)\left(E-E_{d_2}+\frac{i}{2}\Gamma_{d_2}\right)} + \rm{H.~c.}\right)\, . \nonumber
\end{eqnarray}
Using the expressions of the four matrix elements given in Eqs.~(\ref{pneetE1}), (\ref{eler}), (\ref{elnr}), (\ref{augME}), and the summation properties of the  Clebsch-Gordan coefficients, we arrive to the interference cross section term 
\begin{eqnarray}
\label{drinterf}
&&\sigma_{24}(E)=\frac{16\pi^3}{p^2}\rho_iB\!\uparrow\!(E1,I_g\to I_e)(2J_2+1)(2j+1)(-1)^{3I_e+I_g}
 \frac{k^{L+1}}{(2L+1)!!}\sqrt{\frac{L+1}{L(2L+1)^3}}\left(\begin{array}{ccc}J_2 & L& j\\\frac{1}{2}& 0 & -\frac{1}{2}\end{array}\right)^2
\nonumber \\
&&\times R_{L,J_1,J_2}^{(E)}R^{\rm{x-ray}}_{J_2,J_1,L} |\langle \alpha_1 J_1 \|H_{ee}\| [\alpha_i J_i  j ]_{J_1}\rangle|^2
\, \mathrm{Re}\left(\frac{1}{ \left(E-E_{d_1}-\frac{i}{2}\Gamma_{d_1}\right) \left(E-E_{d_1}+\frac{i}{2}\Gamma_{d_1}\right)\left(E-E_{d_2}+\frac{i}{2}\Gamma_{d_2}\right)}\right)\, .
\end{eqnarray}
\end{widetext}
The radial matrix element $R^{\rm{x-ray}}_{J_2,J_1,L}$ corresponding to x-ray emission has the same expression as $R^{RR}_{J_2,j,L}$ in Eq.~(\ref{rrme}) with the difference that 
 both initial and final electronic states are bound. The continuum electron partial wave functions $g_{\varepsilon\kappa}(r)$ and $f_{\varepsilon\kappa}(r)$ are therefore replaced by $g_{n_1\kappa_1}(r)$ and $f_{n_1\kappa_1}(r)$, respectively.

\subsection{NEEC interference term}
Finally, the last interference term of interest is the one that involves in both channels the excitation of the nucleus. Instead of DC, the free electron can recombine into the bound shell with the simultaneous excitation of the nucleus in the process of NEEC. The excited nuclear state decays in both channels radiatively. Thus, the interference occurs between DC followed by NEET on the one hand and NEEC on the other hand. 

The interference term in the cross section is given by
\begin{widetext}
\begin{eqnarray}
&&\sigma_{34}(E)=\frac{(2\pi)^4}{p^2}\rho_i\rho_f\sum_{M,M_f,M_2}\frac{1}{4\pi}\int d\Omega_{\vec{p}}\frac{1}{2}\sum_{m_s}\frac{1}{2J_i+1}\sum_{M_i}\frac{1}{2I_g+1}\sum_{M_g} \nonumber \\
&& \times
\left(\sum_{M_e}\frac{\langle I_g M_f,(E)kLM|H_{nr}|I_e M_e\rangle^*\langle I_e M_e,\alpha_2 J_2 M_2|H_{en}|I_g M_g,\alpha_i J_i M_i\vec{p}m_s\rangle^*}{E-E_{d_2}-\frac{i}{2}\Gamma_{d_2}}
\right. \nonumber \\
&&\times  \left.
\sum_{M'_e}\sum_{\alpha'_1 J'_1 M'_1}\frac{\langle I_g M_f,(E)kLM|H_{nr}|I_e M'_e\rangle \langle I_e M'_e,\alpha_2 J_2 M_2|H_{en}|I_g M_g,\alpha'_1 J'_1 M'_1\rangle \langle \alpha'_1 J'_1 M'_1|H_{ee}|\alpha_i J_i M_i \vec{p} m_s\rangle}
{\left(E-E_{d_1}+\frac{i}{2}\Gamma_{d_1}\right)\left(E-E_{d_2}+\frac{i}{2}\Gamma_{d_2}\right)} + \rm{H.~c.}\right)\, . \nonumber \\
\end{eqnarray}
\end{widetext}
NEEC and NEET are described by the same Hamiltonian $H_{en}$. However, the matrix element for NEEC denotes in the electronic part a transition from a continuum state to a bound state, i.e., recombination, while for NEET we have a transition between bound states. Using the continuum electron wave function expansion in partial waves and after performing the summations we obtain
\begin{eqnarray}
\label{neecinterf}
&&\sigma_{34}(E)=\frac{64\pi^4}{p^2}\rho_iB\!\uparrow\!(E1,I_g\to I_e)B\!\downarrow\!(E1,I_e\to I_g) \nonumber \\
&&\times \frac{k^{2L+1}}{[(2L+1)!!]^2}C\left(J_{2}\ L\ j;\frac{1}{2}\ 0\ \frac{1}{2}\right)^2 R_{L,j,J_2}^{(E)} R_{L,J_1,J_2}^{(E)}
\nonumber \\
&&\times \frac{L+1}{L(2L+1)^2}\frac{
\langle \alpha_1 J_1 \|H_{ee}\| [\alpha_i J_i  j ]_{J_1}\rangle}{(E - E_{d_2})^2 + \frac{\Gamma^2_{d_2}}{4}}
\nonumber \\
&&\times(2J_2+1)\mathrm{Re}\left(\frac{1}{  \left(E-E_{d_1}+\frac{i}{2}\Gamma_{d_1}\right)}\right)\, .
\end{eqnarray}
%

\section{Numerical Results \label{results}}
In the following we present in detail a  numerical example for the process of NEET in HCI. One of the interesting candidates is  $^{237}\mathrm{Np}$
with its 102.959 keV $E1$ nuclear transition which comes close to the $L3\rightarrow K$ atomic transition. The mismatch between the atomic and nuclear transition energies is listed in Ref.~\cite{NEET_list} as 1.89 keV for neutral atoms. This energy difference enters the denominator of the NEET probability in the second power, as showed in Eq.~(\ref{pneet}). Consequently,  a reduction of the mismatch can enhance substantially the NEET probability, up to the maximum of $\Delta E=0$ in which case the denominator in  Eq.~(\ref{pneet}) will be determined by the width of the first intermediate state $\Gamma_{d_1}$, i.e. the width of the electronic state.

The first step is therefore to identify the optimal electronic configuration for NEET and to calculate the transition energies between the $2p$ and $1s$ subshells for different charge states and electronic configurations. For the calculation of bound electron level energies and widths we have used 
 the {\textsc GRASP92} package \cite{Par96}. The acronym  {\textsc GRASP} stands for {\bf G}eneral-purpose {\bf R}elativistic {\bf A}tomic {\bf S}tructure {\bf P}rogram and is a suite of {\textsc FORTRAN} codes for various calculations of relativistic atomic structure. The MCDF approximation is used for the calculation of atomic stationary states and transitions among them \cite{Dyall}. {\textsc GRASP92} is an improvement of the previous versions and includes approximate QED corrections for the electronic energy levels. The finite size of the nucleus, i.e., its radius $R_0$, is also considered in the {\textsc GRASP92} wave functions and has a sensitive effect on the energy levels of the bound inner-shell electron. The accuracy of the {\textsc GRASP92} bound electron energies is expected to be on the level of less than 10 eV for few-electron systems such as He-like and Li-like ions, well within the experimental uncertainties.
The calculated atomic transition energies and widths are listed in Table~\ref{enwid}  for a number of ion configurations. We see that the energy mismatch is smallest for the Li-like configurations. 
\renewcommand\arraystretch{1.3}
\begin{table*}[htb]
\caption{\label{enwid} The electronic transition energy $E_{a}$, mismatch to the nuclear transition energy $\Delta E=E_n - E_a$ and excited electronic state width  $\Gamma_{d_1}$ for various ion configurations as initial state for NEET in $^{237}\mathrm{Np}$.}
\begin{ruledtabular}
\begin{tabular}{lcrc}
Configuration & $E_{a}$ (keV) & $\Delta E$ (keV)  & $\Gamma_{d_1}$ (eV) \\
\hline
$[1s^12s^22p_{1/2}^22p_{3/2}^4]_{1/2+}$	& 101.664	& 	1.295	& 70	\\

$[1s^12s^22p_{1/2}^22p_{3/2}^3]_{1-}$	& 101.701	& 	1.258	& 76	\\

$[1s^12s^12p_{3/2}^1]_{1/2-}$	& 102.879 & 	0.080	& 	30 \\
$[1s^12s^12p_{3/2}^1]_{3/2-}$	& 102.999	& -0.040		& 22	\\

$[1s^12p_{3/2}^1]_{1-}$	& 103.279& 	-0.320	& 	29\\

\end{tabular}

\end{ruledtabular}
\end{table*}

Choosing the most advantageous configurations as the  $[1s2s2p_{3/2}]_J$ set, we proceed to calculate the NEET probability and the resonance strength for the three-step process of DC followed by NEET and $\gamma$-ray emission. The results, including continuum electron energies,  Auger rates, the NEET probability and the resonance strength are presented in Table~\ref{NpNEET}. The Auger rates were obtained with a computer code extension of {\textsc GRASP} developed for DR calculations \cite{Zimmerer2,tr,uranium}. The radial wave functions that enter the integral in Eq.~(\ref{radint}) are obtained with the {\textsc GRASP92} code.
The nuclear radiative rate was calculated
according to the formula \cite{Ring}
\begin{equation}
A_{\gamma}(E1)=\frac{8\pi(L+1)}{L[(2L+1)!!]^2}\frac{E^{2L+1}}{c}B\!\downarrow\!(\lambda L,I_e\to
I_g)\ ,
\end{equation}
with $L=1$. The two reduced transition
probabilities for the emission, respectively the absorption of a $\gamma$ ray are
related through the formula
\begin{equation}
B\!\downarrow\!(\lambda L,I_e\to I_g)=\frac{2I_g+1}{2I_e+1}B\!\uparrow\!(\lambda L,I_g\to I_e)\ .
\end{equation}
The width of the excited nuclear state is then
\begin{equation}
\label{width}
\Gamma_{d_2}=\sum_{\nu}\left(A^{\nu}_{\gamma}+A^{\nu}_{\rm IC}\right) \,
\end{equation}
where $A_{\rm IC}$ is the IC rate for the decay of state $d_2$ and $\nu$ is the sum over all decay channels.
In our case, due to the high binding potential of the two $K$-shell electrons, only IC of the $2s $ electron is possible. The internal conversion coefficient for this state is small, $2.14\times 10^{-2}$ \cite{Roesel} for the 102.959 keV transition, such that the nuclear width is determined by the radiative decay of the nucleus. It should be noted that the 102.959 keV nuclear state does not only  decay directly to the ground state, but also via two intermediate states at $E=59.5409$~keV and $E=33.196$~keV, such that both radiative and IC decay rates sum up three contributions $\nu$. The most important contribution for the nuclear width is given by the radiative and IC decays to the  $59.5409$~keV state.

The calculated NEET probabilities are small, on the order of $10^{-9}$. Compared to previous relativistic calculations that considered NEET occurring in neutral atoms  irradiated by x-rays to generate an inner-shell hole \cite{Tkalya1992}, we see  an enhancement of more than three orders of magnitude of the NEET probability. This is due to the very small energy mismatch $\Delta E=E_n-E_a$ for the case of NEET in HCI. After taking into account the different energy mismatch $\Delta E=E_n-E_a$ in the denominator of $\mathcal{P}_{neet}$ considered in the previous works, we find that our values for the NEET matrix element confirm the results in Ref.~\cite{Tkalya1992,Harston2001} and are three orders of magnitude smaller than the results in Ref.~\cite{Ho1993}.

The cross section as a function of the free electron energy has the shape of a narrow Lorentzian profile, with the width determined by the excited nuclear state width $\Gamma_{d_2}=2.75\times 10^{-6}$~eV and the peak value at the nuclear transition energy $E=$102.959 keV of $\sigma=1.8\times 10^{-4}$~b for the $[1s^12s^12p_{3/2}^1]_{3/2}$ and $\sigma=6.2\times 10^{-5}$~b for $[1s^12s^12p_{3/2}^1]_{1/2}$ initial configurations, respectively.

\begin{table*}[htb]
\caption{\label{NpNEET} NEET probabilities and total resonance strengths for the three-step process of DC into the $[1s2s2p_{3/2}]_J$ doubly-excited state followed by NEET and $\gamma$ decay of the nuclear excited state. The atomic transition energy $E_{a}$, continuum electron energy  $E_{\vec{p}}$ and intermediate state widths are given, together with the Auger rate $A_{Au}$. }
\begin{ruledtabular}
\begin{tabular}{lrrccclr}
J & $E_{a}$ (keV) & $E_{\vec{p}}$ (keV)  & $\Gamma_{d_1}$ (eV) & $\Gamma_{d_2}$ (eV) & $A_{Au}$ (1/s) & $\mathcal{P}_{neet}$ & $\int \sigma(E) dE$ (b$\,$eV) \\
\hline
1/2	& 102.879	& 69.353	& 30	&2.75$\times 10^{-6}$	&	3.36$\times 10^{12}$& 7.2$\times 10^{-10}$	& 	8.19$\times 10^{-8}$\\
3/2	& 102.999	& 69.275	& 22	&2.75$\times 10^{-6}$	&	2.52$\times 10^{12}$&	2.8$\times 10^{-9}$& 	2.36$\times 10^{-7}$\\
\end{tabular}

\end{ruledtabular}
\end{table*}

After investigating the magnitude of the NEET cross section, we turn now to the competing processes denoted in Section \ref{interf} as channels (1), (2) and (3) and the interference terms between them.
For $^{237}\mathrm{Np}$, the nuclear transition is electric dipole allowed, such that the competing channels of RR (direct recombination into the $2s$ orbital) and DR (DC in the $[1s^12s^12p_{3/2}^1]_{J}$ state followed by x-ray decay) are strong. 
The total cross sections for channels (1) and (3) were calculated following the approaches presented in Refs.~\cite{Eichler2000,palffy2006,interference2007}.
For the NEEC,  RR  and x-ray rates we have evaluated the radial integrals $R_{L,j,J_2}^{(E)}$,
$R^{RR}_{J_2,j,L}$ and $R^{\rm{x-ray}}_{J_2,J_1,L}$ that enter Eqs.~(\ref{rrinterf}), (\ref{drinterf}) and (\ref{neecinterf})    numerically. Relativistic Coulomb-Dirac wave functions for the continuum electron and {\textsc GRASP92} bound electron wave functions were used. 

Due to a rather small nuclear reduced transition probability $B(E1)$ \cite{NuclearDataBase}, the NEEC resonance strength for electronic capture into the $2s$ orbital with the simultaneous excitation of the 102.959 keV nuclear state is not very high, only $S=6.21\times 10^{-4}$~b$\,$eV.  Just as for the case of RR, in the calculation of the NEEC rate we have considered only the recombination of the continuum electron with the partial wave ($\kappa=1$ for the configuration with $J=1/2$ and $\kappa=-2$ for the configuration with $J=3/2$, respectively) that contributes to the interference.  
The NEEC resonance strength is still  three to four orders of magnitude larger than the NEET resonance strengths presented in Table~\ref{NpNEET}. This can be traced back to the 
overlap between the bound and continuum radial wave functions for $\mathrm{Np}^{91+}$.  We have investigated and compared the magnitude of the radial integrals $R_{L,j,J_2}^{(E)}$ and $R_{L,J_1,J_2}^{(E)}$ for several values of the continuum electron energies. It turns out that for HCI, where the continuum electron recombines into the strong field of the ion, the radial integrals involving the continuum electron wavefunctions are larger than the ones involving only bound electronic states. Among NEET bound-bound radial integrals, we find that the overlap is largest for $2s\rightarrow 1s$ transitions.

The energy dependence of the cross sections for the four recombination channels is decisive for the behavior of the interference terms. Both channels (3) and (4), which involve the excitation of the nucleus, present a very narrow Lorentzian energy  profile centered on the nuclear transition energy. Although the resonance strength, i.e., the integrated total cross section is very small, the narrow width of the Lorentzian profile (determined by the nuclear excited state width) makes the cross section at resonance to be comparatively large: $10^{-4}$ and $10^{-5}$ b for the three-step process involving NEET and 143 b for the two-step process involving NEEC. 

The atomic recombination processes, on the other hand, do not present such narrow resonances. RR is a direct process with no resonance profile and for the energy interval of interest around the nuclear transition energy the cross section can be approximated as constant at the value 27.6 b. For the total RR cross section, two partial waves in the continuum electron expansion contribute to the cross section. However, only one of these partial waves can participate in the interference.  The contributions of the two partial waves in the total cross section are 6.83 b for  $\kappa=1$ (capture into $[1s^12s^12p_{3/2}^1]_{1/2}$) and  21.8 b for  $\kappa=-2$ (capture into $[1s^12s^12p_{3/2}^1]_{3/2}$).

\begin{figure}[h]
  \centering
  \includegraphics*[width=8.7cm]
             {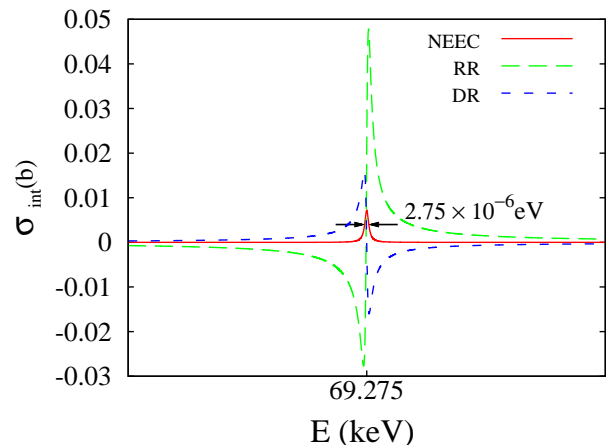}
  \caption{ Interference terms in the cross section $\sigma_{14}$ (RR), $\sigma_{24}$ (DR) and $\sigma_{34}$ (NEEC) as a function of the continuum electron energy for recombination into $\mathrm{Np}^{91+}$. See text for further explanations.}
  \label{plot_interf}
\end{figure}
\begin{figure}[h]
  \centering
  \includegraphics*[width=8.7cm]
             {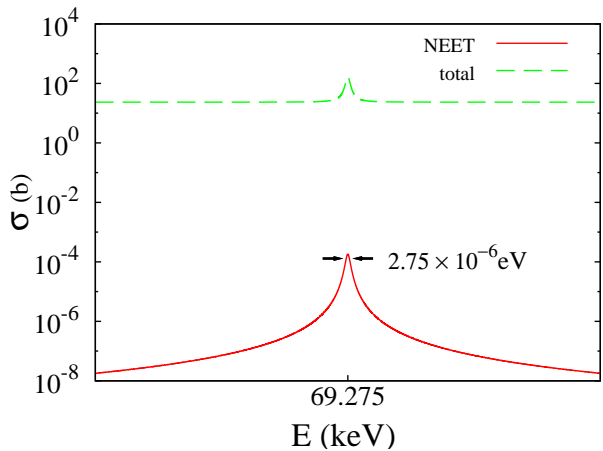}
  \caption{Total cross section and $\sigma_4(E)$ (NEET) for the recombination of a free electron into $\mathrm{Np}^{91+}$ at the resonance energy for DC and NEEC. See text for further explanations. }
  \label{total_cs}
\end{figure}

DR is a resonant process whose  width of the Lorentzian profile is determined by the width of the excited electronic state, on the order of tens of eV.  Consequently the resonance cross section is only 2.44~b and 2.38~b for the $[1s^12s^12p_{3/2}^1]_{3/2}$ and $[1s^12s^12p_{3/2}^1]_{1/2}$ configurations, respectively, at the continuum electron energy corresponding to the nuclear resonance, and stays constant on the whole narrow energy region of interest. Thus, although the total resonance strength of both RR and DR is orders of magnitude larger than the ones of the processes involving nuclear excitation, their values at the nuclear resonance energies are not very large. For the case of 
$^{237}\mathrm{Np}$, the RR and DR cross sections at the resonance energy are orders of magnitude larger than the one of NEET, but much smaller than the one for NEEC.
 The interference terms $\sigma_{14}(E)$, $\sigma_{24}(E)$ and $\sigma_{34}(E)$ 
all inherit the very narrow width of channel (4), and the magnitude of the interference cross section is determined by the resonance cross sections of the individual processes. The RR and DR contributions in the interference are expected to be largest for electric dipole allowed transitions, as it is the case of $^{237}\mathrm{Np}$. In Fig. \ref{plot_interf} we plot the three interference terms considering for DC the $[1s^12s^12p_{3/2}^1]_{3/2-}$ configuration as a function of the continuum electron energy on a very narrow interval around the resonance energy. We see that both the purely atomic processes as well as NEEC have interference terms with similar peak values at the resonance energy, on the order of $10^{-2}$~b. This value, although small, is two to three orders of magnitude larger than the NEET cross section value. We note that this situation is qualitatively different than the one presented in Ref.~\cite{interference2007} for the interference between RR and NEEC. In that case, compared to the large peak value of the NEEC cross section, the interference term was orders of magnitude smaller and therefore negligible.

The total cross section of process (4) with and without including the interference terms and the competing processes are presented in Fig. \ref{total_cs} for DC leading to the $[1s^12s^12p_{3/2}^1]_{3/2-}$ state. The cross section plateau  is determined by the RR and DR cross sections, while the peak value at the  resonance energy is determined by NEEC.
 Surprisingly, although RR and DR have much larger resonance strengths than the processes involving nuclear transitions, it is  NEEC which helps rise the total cross section of the three-step NEET process at the resonance energy. Due to the very large resonance cross section of channel (3), we see that among the interference terms,  the  term $\sigma_{34}(E)$ is smaller but comparable with $\sigma_{14}(E)$ and $\sigma_{24}(E)$. However, more than the interference cross section term, it is the NEEC term itself that is the relevant contribution for nuclear excitation  in the recombination process. 
We conclude that in HCI, the two nuclear excitation mechanisms NEET and NEEC, when both possible, might deliver contributions of similar importance and  should be treated on equal footing.


\section{\label{sum} Summary}

In this work we have investigated a new aspect of electron recombination into HCI
involving the coupling of the atomic shell to the nucleus in the process of NEET.
Our scenario involves the resonant process of DC into HCI to create the electronic hole needed for NEET.
HCI present the advantage that the atomic level energies are very sensitive to the ion
charge state and offer the possibility to optimize the match between atomic and nuclear 
transition energies. The NEET probability can thus be enhanced by several orders of 
magnitude compared to neutral atoms.
This can be of great advantage for experiments aiming at 
investigating NEET. Furthermore, both HCI and DC are predominant in dense astrophysical 
plasmas where NEET is expected to play an important role for the population
of nuclear excited states.

We have developed  a versatile formalism for describing complex
processes actively involving atomic electrons and nuclei and derived  total
cross sections and resonance strengths for the direct and resonant channels of electron recombination.
The total and interference cross section terms for the processes of RR, DR, DC followed by NEET and $\gamma$ decay and NEEC followed by $\gamma$ 
decay were deduced and their magnitude investigated for a test case. Our results show that for HCI
NEEC may be the most important contribution to the total recombination cross sections and that the interference terms, although small, are still larger than the NEET  cross section.  This is qualitatively different from case of  the typical NEET scenario investigated until now in experiments, where the specific inner shell vacancy is created in neutral gold atoms $^{197}\mathrm{Au}$ \cite{Kishimoto2000} and NEEC  cannot occur simultaneously with NEET.  
The impact of our findings for nuclei of astrophysical relevance in
isomer depletion and nucleosynthesis is still to be investigated. Calculations for such particular cases as well as 
estimates about the feasibility of a new type of  NEET experiment with HCI are in progress.


\begin{acknowledgments}

The authors would like to thank A. Lapierre for bringing this problem into their attention,
 Z. Harman for the computation of the Auger rates and C. H. Keitel for fruitful discussions. SKA acknowledges the hospitality of
the Theory Division at the Max-Planck-Institut f\"ur Kernphysik.

\end{acknowledgments}

\bibliography{palffy}

\end{document}